\documentclass[aps,preprint,nofootinbib,superscriptaddress]{revtex4}%
\usepackage{amssymb}
\usepackage{amsmath}
\usepackage{epsfig}
\usepackage{amsfonts}
\usepackage{graphicx}
\usepackage{hyperref}%
\providecommand{\U}[1]{\protect\rule{.1in}{.1in}}
\begin{document}
\title{Yang-Mills Instanton Sheaves with Higher Topological Charges}
\author{Sheng-Hong Lai}
\email{xgcj944137@gmail.com}
\affiliation{Department of Electrophysics, National Chiao-Tung University, Hsinchu, Taiwan, R.O.C.}
\author{Jen-Chi Lee}
\email{jcclee@cc.nctu.edu.tw}
\affiliation{Department of Electrophysics, National Chiao-Tung University, Hsinchu, Taiwan, R.O.C.}
\author{I-Hsun Tsai}
\email{ihtsai@math.ntu.edu.tw}
\affiliation{Department of Mathematics, National Taiwan University, Taipei, Taiwan, R.O.C. }
\date{\today}

\begin{abstract}
We explicitly construct $SL(2,C)$ (or $SU(2)$ complex) Yang-Mills (weakly)
three and four instanton sheaves on $CP^{3}$. These results extend the
previous construction of Yang-Mills (weakly) instanton sheaves with
topological charge two \cite{Ann2}.

\end{abstract}
\maketitle
\tableofcontents
%

\setcounter{equation}{0}
\renewcommand{\theequation}{\arabic{section}.\arabic{equation}}%

\section{Introduction}

One of the most important developments for the interplay of quantum field
theory and algebraic geometry in 1970s was the discovery of Yang-Mills (YM)
instantons.\bigskip\ Historically, the first BPST $SU(2)$ $1$-instanton
solution \cite{BPST} with $5$ moduli parameters was discovered in 1975. Soon
later the CFTW $k$-instanton solutions \cite{CFTW} with $5k$ moduli parameters
were constructed, and then the so-called JNR $k$-instanton solutions with
$5k+4$ ($5$,$13$ for $k=1$,$2$) \cite{JR} parameters was proposed based on the
$4D$ conformal symmetry group of YM equation. This important issue was finally
solved in 1978 by ADHM \cite{ADHM} using method in algebraic geometry. The
complete solutions of finite action (anti)self-dual YM (SDYM) equation was
found to contain $8k-3$ moduli parameters for each $k$th homotopy class.
Applications of YM instantons on quantum field theory, in particular QCD, can
be found at \cite{U(1),the}

The original ADHM theory used the monad construction combining with the
Penrose-Ward transform to construct the most general instanton solutions by
establishing an one to one correspondence between (anti)self-dual
$SU(2)$-connections on $S^{4}$ and global holomorphic vector bundles of rank
two on $CP^{3}$. The latter one was much easier to identify and the explicit
closed forms of the complete $SU(2)$ instanton solutions for $k\leq3$ had been
worked out in \cite{CSW}. From physicist point of view, the YM instantons can
also be seen through the lens of string theory. They can be embedded in
Heterotic stringy soliton solutions \cite{andy,andy2}. They can also be
described in terms of worldsheet supersymmetric sigma-models \cite{ed}. On the
other hand, one simple way to see ADHM instantons from physicist point of view
was to use the brane constructions in type II string theory \cite{D1,D2,D3}.

In a recent paper \cite{Ann}, the quaternion calculation of $SU(2)$ ADHM
construction was generalized to the biquaternion calculation with
biconjugation operation, and a class of non-compact $SL(2,C)$ (or $SU(2)$
complex) YM instanton solutions with $16k-6$ parameters for each $k$th
homotopy class was constructed. The number of parameters $16k-6$ was first
conjectured by Frenkel and Jardim in \cite{math2} and was proved recently in
\cite{math3} from the mathematical point of view. These new $SL(2,C)$
instanton solutions include as a subset the previous $SL(2,C)$ $(M,N)$
instantons constructed in 1984 \cite{Lee}.

One important motivation to study $SL(2,C)$ instanton solutions has been to
understand, in addition to the holomorphic vector bundles on $CP^{3}$ in the
ADHM construction which has been well studied in the $SU(2)$ instantons, the
(weakly) instanton sheaves on the projective space. One key hint for the
existence of $SL(2,C)$ (weakly) instanton sheaves on $CP^{3}$ was the
discovery \cite{Ann} of singularities for $SL(2,C)$ instanton solutions on
$S^{4}$ which can not be gauged away as in the case of $SU(2)$ instantons.

The first YM (weakly) instanton sheaves was constructed recently only for
$SL(2,C)$ $2$-instanton solutions \cite{Ann2}. It is thus of interest to see
whether there exist general YM (weakly) $k$-instanton sheaves with higher
topological charges. Since it was shown that \cite{Ann2} the $SL(2,C)$
extended $(M,N)$ $k$-instanton solutions with $10k$ parameters on $S^{4}$
correspond to the locally free sheaves or holomorphic vector bundles on
$CP^{3}$, one needs to consider the non-diagonal $k$-instanton solutions with
$k\geq3$ in order to get YM (weakly) instanton sheaves.

In this paper, we will explicitly construct $SL(2,C)$ Yang-Mills (weakly)
three and four instanton sheaves on $CP^{3}$ which extend our previous
construction of (weakly) two instanton sheaves. For the case of (weakly) three
instanton sheaves, we make use of a set of $SU(2)$ ADHM three instanton data
to do the construction. For the case of four (weakly) instanton sheaves, in
addition to the known explicit $SU(2)$ $k$-instanton solutions with $k\leq3$,
there existed in the literature the so-called $SU(2)$ ADHM \textit{symmetric}
four instanton \cite{4in} solutions. The motivation to explicitly calculate
these ADHM instanton solutions with higher topological charges was to
construct the approximate minimum energy skyrmion fields in the Skyrme model
of low energy hadronic physics.

Indeed, it was suggested by Atiyah and Manton \cite{AM} that low energy
skymion fields of charges $k$ can be approximated by computing holonomy of
$k$-instanton on $R^{4}$ along lines parallel to the Euclidean time direction
\cite{MS}. In constrast to the Skymion fields construction, for our purpose in
this paper instead, our motivation is to use these $SU(2)$ ADHM instanton data
to construct $SL(2,C)$ (weakly) instanton sheaves on $CP^{3}$ with higher
topological charges. Surprisingly, we will see that there do exist (weakly)
instanton sheaf structures on these symmetric four instanton solutions. It is
thus an interesting issue to understand the relationship between YM
\textit{symmetric} instantons \cite{3in,4in,7in} on $S^{4}$ and YM (weakly)
instanton \textit{sheaves} on $CP^{3}$ constructed in this paper. (The idea of
"weakly instanton sheaves" and "instanton sheaves" will be explained in the
beginning of subsection III. B).

This paper is organized as following. In section II, we review the
biquaternion construction of $SL(2,C)$ YM instanton which was developed
recently by the present authors \cite{Ann}. We also introduce the complex ADHM
equations \cite{Donald} and the monad construction which will be used in the
follow-up sections. In section III, we discuss in details the construction of
a class of $SL(2,C)$ YM (weakly) three instanton sheaves. The construction was
extended to the $SL(2,C)$ YM (weakly) four instanton sheaves by using a class
of $SU(2)$ symmetric four instanton solutions in section IV. A brief
conclusion was given in section V.

\section{Biquaternions and $SL(2,C)$ Yang-Mills Instanton Construction}

We first briefly review the $SL(2,C)$ YM theory. There are two linearly
independent choices of $SL(2,C)$ group metric \cite{WY}
\begin{equation}
g^{a}=%
\begin{pmatrix}
I & 0\\
0 & -I
\end{pmatrix}
,g^{b}=%
\begin{pmatrix}
0 & I\\
I & 0
\end{pmatrix}
\end{equation}
where $I$ is the $3\times3$ unit matrix. In general, one can choose
\begin{equation}
g=\cos\theta g^{a}+\sin\theta g^{b}%
\end{equation}
where $\theta$ is a real constant. It can be shown that $SL(2,C)$ is
isomorphic to $S^{3}\times R^{3},$ and one can calculate its third homotopy
group \cite{Lee}%
\begin{equation}
\pi_{3}[SL(2,C)]=\pi_{3}[S^{3}\times R^{3}]=\pi_{3}(S^{3})\cdot\pi_{3}%
(R^{3})=Z\cdot I=Z\newline\newline%
\end{equation}
\newline where $I$ is the identity group, and $Z$ is the integer group.

Wu and Yang \cite{WY} have shown that complex $SU(2)$ gauge fields are related
to the real $SL(2,C)$ gauge fields. Starting from $SU(2)$ complex gauge field
formalism, one can write down the $SL(2,C)$ YM equations. For the complex
gauge field
\begin{equation}
G_{\mu}^{a}=A_{\mu}^{a}+iB_{\mu}^{a},
\end{equation}
the corresponding complex field strength is defined as ($g=1$)
\begin{equation}
F_{\mu\nu}^{a}\equiv H_{\mu\nu}^{a}+iM_{\mu\nu}^{a},a,b,c=1,2,3
\end{equation}
where
\begin{align}
H_{\mu\nu}^{a}  &  =\partial_{\mu}A_{\nu}^{a}-\partial_{\nu}A_{\mu}%
^{a}+\epsilon^{abc}(A_{\mu}^{b}A_{\nu}^{c}-B_{\mu}^{b}B_{\nu}^{c}),\nonumber\\
M_{\mu\nu}^{a}  &  =\partial_{\mu}B_{\nu}^{a}-\partial_{\nu}B_{\mu}%
^{a}+\epsilon^{abc}(A_{\mu}^{b}B_{\nu}^{c}-A_{\mu}^{b}B_{\nu}^{c}).
\end{align}
The $SL(2,C)$ YM equation can then be written as
\begin{align}
\partial_{\mu}H_{\mu\nu}^{a}+\epsilon^{abc}(A_{\mu}^{b}H_{\mu\nu}^{c}-B_{\mu
}^{b}M_{\mu\nu}^{c})  &  =0,\nonumber\\
\partial_{\mu}M_{\mu\nu}^{a}+\epsilon^{abc}(A_{\mu}^{b}M_{\mu\nu}^{c}-B_{\mu
}^{b}H_{\mu\nu}^{c})  &  =0,
\end{align}
and the $SL(2,C)$ SDYM equations are%
\begin{align}
H_{\mu\nu}^{a}  &  =\frac{1}{2}\epsilon_{\mu\nu\alpha\beta}H_{\alpha\beta
},\nonumber\\
M_{\mu\nu}^{a}  &  =\frac{1}{2}\epsilon_{\mu\nu\alpha\beta}M_{\alpha\beta}.
\label{self}%
\end{align}
YM equation for the choice $\theta=0$ in this paper can be derived from the
following Lagrangian
\begin{equation}
L=\frac{1}{4}(H_{\mu\nu}^{a}H_{\mu\nu}^{a}-M_{\mu\nu}^{a}M_{\mu\nu}^{a}).
\end{equation}

\bigskip We now proceed to review the construction of $SL(2,C)$ YM instantons
\cite{Lee,Ann}. We will use the convention $\mu=0,1,2,3$ and $\epsilon
_{0123}=1$ for $4D$ Euclidean space. In contrast to the quaternion in the
$Sp(1)$ ($=SU(2)$) ADHM construction, the authors of \cite{Ann} used
\textit{biquaternion} to construct $SL(2,C)$ YM instantons. A quaternion $x$
can be written as%
\begin{equation}
x=x_{\mu}e_{\mu}\text{, \ }x_{\mu}\in R\text{, \ }e_{0}=1,e_{1}=i,e_{2}%
=j,e_{3}=k \label{x}%
\end{equation}
where $e_{1},e_{2}$ and $e_{3}$ anticommute and obey%
\begin{align}
e_{i}\cdot e_{j}  &  =-e_{j}\cdot e_{i}=\epsilon_{ijk}e_{k};\text{
\ }i,j,k=1,2,3,\\
e_{1}^{2}  &  =-1,e_{2}^{2}=-1,e_{3}^{2}=-1.
\end{align}
The conjugate quaternion is defined to be%

\begin{equation}
x^{\dagger}=x_{0}e_{0}-x_{1}e_{1}-x_{2}e_{2}-x_{3}e_{3}%
\end{equation}
so that the norm square of a quaternion is%
\begin{equation}
|x|^{2}=x^{\dagger}x=x_{0}^{2}+x_{1}^{2}+x_{2}^{2}+x_{3}^{2}. \label{norm}%
\end{equation}
Occasionaly the unit quaternions can be expressed as Pauli matrices%
\begin{equation}
e_{0}\rightarrow%
\begin{pmatrix}
1 & 0\\
0 & 1
\end{pmatrix}
,e_{i}\rightarrow-i\sigma_{i}\ \text{; }i=1,2,3. \label{pauli}%
\end{equation}

A biquaternion (or complex-quaternion) $z$ can be written as%
\begin{equation}
z=z_{\mu}e_{\mu}\text{, \ }z_{\mu}\in C
\end{equation}
or%
\begin{equation}
z=x+yi
\end{equation}
where $x$ and $y$ are quaternions and $i=\sqrt{-1}$. In the recent
construction of $SL(2,C)$ YM instantons \cite{Ann}, the biconjugation
\cite{Ham} of $z$ is defined to be%
\begin{equation}
z^{\circledast}=z_{\mu}e_{\mu}^{\dagger}=z_{0}e_{0}-z_{1}e_{1}-z_{2}%
e_{2}-z_{3}e_{3}=x^{\dagger}+y^{\dagger}i,
\end{equation}
in contrast to the complex conjugation%
\begin{equation}
z^{\ast}=z_{\mu}^{\ast}e_{\mu}=z_{0}^{\ast}e_{0}+z_{1}^{\ast}e_{1}+z_{2}%
^{\ast}e_{2}+z_{3}^{\ast}e_{3}=x-yi.
\end{equation}
The norm square of a biquaternion is defined to be%
\begin{equation}
|z|_{c}^{2}=z^{\circledast}z=(z_{0})^{2}+(z_{1})^{2}+(z_{2})^{2}+(z_{3})^{2},
\end{equation}
which is a \textit{complex} number in general and a subscript $c$ is used in
the norm.

We now briefly review how to extend the quaternion construction of ADHM
$SU(2)$ instantons to the $SL(2,C)$ YM instantons. The first step was to
introduce the $(k+1)\times k$ biquaternion matrix%
\begin{equation}
\Delta(x)=A+Dx,
\end{equation}
which satisfies the quadratic condition%
\begin{equation}
\Delta(x)^{\circledast}\Delta(x)=f^{-1}=\text{symmetric, non-singular }k\times
k\text{ matrix for }x\notin J. \label{JJJ}%
\end{equation}

Note that for $x\in J,$ $\det\Delta(x)^{\circledast}\Delta(x)=0$. The set $J$
is called singular locus or jumping lines in the mathematical literature.
There are no jumping lines for the case of $SU(2)$ instantons on $S^{4}$. On
the other hand, in the $SU(2)$ quaternion case, the symmetric condition on
$f^{-1}$ implies $f^{-1}$ is real; while for the $SL(2,C)$ biquaternion case,
it implies $f^{-1}$ is \textit{complex} which means $[\Delta(x)^{\circledast
}\Delta(x)]_{ij}^{\mu}=0$ for $\mu=1,2,3.$

To construct the self-dual gauge fields, one introduces a $(k+1)\times1$
dimensional biquaternion vector $v(x)$ satisfying the two conditions%
\begin{subequations}
\begin{align}
v^{\circledast}(x)\Delta(x)  &  =0,\label{null}\\
v^{\circledast}(x)v(x)  &  =1. \label{norm2}%
\end{align}
Finally one can calculate the gauge fields%

\end{subequations}
\begin{equation}
G_{\mu}(x)=v^{\circledast}(x)\partial_{\mu}v(x), \label{A}%
\end{equation}
which is a $1\times1$ biquaternion.

In the canonical form of the construction, one can set%
\begin{equation}
D=%
\begin{bmatrix}
0_{1\times k}\\
I_{k\times k}%
\end{bmatrix}
,A=%
\begin{bmatrix}
\lambda_{1\times k}\\
-y_{k\times k}%
\end{bmatrix}
\label{ab2}%
\end{equation}
where $\lambda$ and $y$ are biquaternion matrices with orders $1\times k$ and
\ $k\times k$ respectively, and $y$ is symmetric $y=y^{T}$. One can show that
in the canonical form the constraints for the moduli parameters become
\cite{Ann}%
\begin{equation}
A_{ci}^{\circledast}A_{cj}=0,i\neq j,\text{ and \ }y_{ij}=y_{ji}. \label{dof}%
\end{equation}
The total number of moduli parameters for $k$-instanton is $16k-6.$ Note that
for the $SU(2)$ instantons, $\lambda$ and $y$ are the usual quaternion matrices.

There was another approach to solve $SL(2,C)$ YM instantons through the
complex ADHM equations \cite{Donald}
\begin{subequations}
\begin{align}
\left[  B_{11},B_{12}\right]  +I_{1}J_{1}  &  =0,\label{adhm1}\\
\left[  B_{21},B_{22}\right]  +I_{2}J_{2}  &  =0,\label{adhm2}\\
\left[  B_{11},B_{22}\right]  +\left[  B_{21},B_{12}\right]  +I_{1}J_{2}%
+I_{2}J_{1}  &  =0 \label{adhm3}%
\end{align}
where $B_{ij}$ are $k\times k$ complex matrices and $J_{i}$ are $2\times k$
complex matrices. They are the \textit{complex ADHM data }$(B_{lm},I_{m,}%
J_{m})$ with $l,m=1,2$. For the case of $SU(2)$ ADHM instantons, we impose the
conditions \cite{Ann2}%
\end{subequations}
\begin{subequations}
\begin{align}
I_{1}  &  =J^{\dagger},I_{2}=-I,J_{1}=I^{\dagger},J_{2}=J,\nonumber\\
B_{11}  &  =B_{2}^{\dagger},B_{12}=B_{1}^{\dagger},B_{21}=-B_{1},B_{22}=B_{2}%
\end{align}
to recover the real ADHM equations%
\end{subequations}
\begin{subequations}
\begin{align}
\left[  B_{1},B_{2}\right]  +IJ  &  =0,\\
\left[  B_{1},B_{1}^{\dagger}\right]  +\left[  B_{2},B_{2}^{\dagger}\right]
+II^{\dagger}-J^{\dagger}J  &  =0.
\end{align}

Indeed one can identify the ADHM data $(B_{lm},I_{m,}J_{m})$ from the moduli
parameters in Eq.(\ref{ab2}) \cite{Ann2}. The first step is to use
Eq.(\ref{pauli}) to transform the biquaternion $A$ in Eq.(\ref{ab2}) into the
explicit matrix representation (EMR). For the $k$-instanton case, the EMR
of\ the $(k+1)\times k$ biquaternion matrix $A$ in Eq.(\ref{ab2}) can be
written as a $2(k+1)\times2k$ complex matrix $A_{E}$\bigskip. The next step is
to use the following \textit{rearrangement rule }for an element $z_{ij}$ in
$A_{E}$ \cite{Ann2}\textit{ }
\end{subequations}
\begin{align}
z_{2n-1,2m-1}  &  \rightarrow z_{n,m}\text{ },\nonumber\\
z_{2n-1,2m}  &  \rightarrow z_{n,k+m}\text{ },\nonumber\\
z_{2n,2m-1}  &  \rightarrow z_{k+n,m}\text{ },\nonumber\\
z_{2n,2m}  &  \rightarrow z_{k+n,k+m}.
\end{align}
to obtain $A_{E}^{r}$. Finally one can then do the following identification
for the complex ADHM data
\begin{equation}
A_{E}^{r}=%
\begin{bmatrix}
J_{1} & J_{2}\\
B_{11} & B_{21}\\
B_{12} & B_{22}%
\end{bmatrix}
.
\end{equation}
Similar procedure can be perform on $A^{\circledast}$. With these
identifications, one can show that the $SL(2,C)$ YM instantons constructed
previously in \cite{Ann2} are solutions of the complex ADHM equations in
Eq.(\ref{adhm1}) to Eq.(\ref{adhm3}).

Finally, in the monad construction of holomorphic vector bundles on the
projective space, one introduces the $\bigskip\alpha\,$and $\beta$ matrices as
functions of homogeneous coordinates $[x:y:z:w]$ of $CP^{3}$ and define%
\begin{subequations}
\begin{align}
\alpha &  =%
\begin{bmatrix}
zB_{11}+wB_{21}+x\\
zB_{12}+wB_{22}+y\\
zJ_{1}+wJ_{2}%
\end{bmatrix}
,\\
\beta &  =%
\begin{bmatrix}
-zB_{12}-wB_{22}-y & zB_{11}+wB_{21}+x & zI_{1}+wI_{2}%
\end{bmatrix}
.
\end{align}
It can be shown that the condition \cite{Ann2}%
\end{subequations}
\begin{equation}
\beta\alpha=0 \label{ker}%
\end{equation}
is satisfied if and only if the complex ADHM equations in Eq.(\ref{adhm1}) to
Eq.(\ref{adhm3}) holds.

In this construction, Eq.(\ref{ker}) implies Im $\alpha$ is a subspace of Ker
$\beta$ which allows one to consider the quotient vector space Ker $\beta$/ Im
$\alpha$ at each point of $CP^{3}$. For the $SU(2)$ ADHM instantons, the map
$\beta$ is surjective and the map $\alpha$ is injective and $\dim($Ker $\beta
$/ Im $\alpha)=k+2-k=2$ on every points of $CP^{3}$, thus one can use the
holomorphic vector bundles of rank $2$ to describe $SU(2)$ instantons.

For the case of $SL(2,C)$ instantons, we will show that for some general
$k$-instantons $\alpha$ may not be injective at some points of $CP^{3}$ and
$\beta$ may or may not be surjective at some other points of $CP^{3}$ for some
ADHM data, so the dimension of the vector space $($Ker $\beta$/ Im $\alpha)$
may vary from point to point on $CP^{3}$, and one is led to use sheaf
description for these $SL(2,C)$ YM instantons or (weakly) "instanton sheaves"
on $CP^{3}$\cite{math2}.

In a recent publication, some $SL(2,C)$ (weakly) $2$-instanton sheaves were
constructed in \cite{Ann2}. In the following sections, we will calculate a
class of Yang-Mills $SL(2,C)$ (weakly) $k$-instanton sheaves with $k=3$ and
$k=4.$

\section{The $SL(2,C)$ (Weakly) Three Instanton Sheaves}

There existed complete construction of the $SU(2)$ $k$-instanton solutions for
$k\leq3$ in the literature. For the higher instanton solutions, there were the
so-called $SU(2)$ ADHM symmetric $4$-instanton \cite{4in} and $7$-instanton
\cite{7in} solutions. The motivation to explicitly calculate these ADHM
symmetric higher instanton solutions was to construct the approximate minimum
energy skyrmion fields in the Skyrme model of low energy hadronic physics. Our
motivation in this paper is to use these $SU(2)$ ADHM data to construct
$SL(2,C)$ (weakly) instanton sheaves with higher topological charges.

We begin with a $SU(2)$ $3$-instanton solution proposed in \cite{3in}%
\begin{equation}
A=%
\begin{bmatrix}
e_{1} & e_{2} & e_{3}\\
0 & e_{3} & e_{2}\\
e_{3} & 0 & e_{1}\\
e_{2} & e_{1} & 0
\end{bmatrix}
, \label{33A}%
\end{equation}
which was tetrahedrally symmetric and was used to calculate the approximate
$3$-Skyrme field. There was a two parameter generalization of the ADHM data
$A$ in Eq.(\ref{33A}) . For our purpose here, we drop out the symmetric
constraints and propose%
\begin{equation}
A_{3}=%
\begin{bmatrix}
ae_{1} & be_{2} & ce_{3}\\
0 & ce_{3} & be_{2}\\
ce_{3} & 0 & ae_{1}\\
be_{2} & ae_{1} & 0
\end{bmatrix}
. \label{3A}%
\end{equation}
\bigskip One can easily calculate%
\begin{equation}
A_{3}^{\circledast}A_{3}=%
\begin{bmatrix}
\left(  a^{2}+b^{2}+c^{2}\right)  e_{0} & 0 & 0\\
0 & \left(  a^{2}+b^{2}+c^{2}\right)  e_{0} & 0\\
0 & 0 & \left(  a^{2}+b^{2}+c^{2}\right)  e_{0}%
\end{bmatrix}
.
\end{equation}
Thus $A_{3}$ proposed in Eq.(\ref{3A}) satisfies the constraints in
Eq.(\ref{dof}) for arbitrary $a,b\ $and $c$ $\in C$ for the $SL(2,C)$ case,
and represents a class of ADHM data. Note that for $a,b\ $and $c$ $\in R$,
$A_{3}$ represents a class of $SU(2)$ $3$-instanton solutions.

We are now ready to check whether there exists instanton sheaf structure for
these $3$-instanton solutions. \bigskip We first calculate $A_{3E}$ , the EMR
of $A_{3}$, and then do the rearrangement rule to obtain $A_{3E}^{r}$
\cite{Ann2} and finally identify the corresponding ADHM data $(B_{lm}%
,I_{m,}J_{m})$ to be%
\begin{align}
J_{1}  &  =%
\begin{bmatrix}
0 & 0 & -ic\\
-ia & b & 0
\end{bmatrix}
,J_{2}=%
\begin{bmatrix}
-ia & -b & 0\\
0 & 0 & ic
\end{bmatrix}
,\\
I_{1}  &  =%
\begin{pmatrix}
ia & 0\\
-b & 0\\
0 & -ic
\end{pmatrix}
,I_{2}=%
\begin{pmatrix}
0 & -ia\\
0 & -b\\
-ic & 0
\end{pmatrix}
,\\
B_{11}  &  =%
\begin{bmatrix}
0 & -ic & 0\\
-ic & 0 & 0\\
0 & 0 & 0
\end{bmatrix}
,B_{21}=%
\begin{bmatrix}
0 & 0 & -b\\
0 & 0 & -ia\\
-b & -ia & 0
\end{bmatrix}
,\\
B_{12}  &  =%
\begin{bmatrix}
0 & 0 & b\\
0 & 0 & -ia\\
b & -ia & 0
\end{bmatrix}
,B_{22}=%
\begin{bmatrix}
0 & ic & 0\\
ic & 0 & 0\\
0 & 0 & 0
\end{bmatrix}
.
\end{align}

\subsection{The $\alpha$ matrix and costable conditions}

The next step is to check the costability conditions \cite{math2}. Note that
in \cite{LLT4} a duality symmetry among stability conditions and costability
conditions for YM instanton sheaf solutions was pointed out with application
to the known sheaf solutions. We will give one example of duality
transformation in subsection C. The result of the duality symmetry first
presented in \cite{LLT4} was to replace the misdeading result of the proof of
equivalence of stability conditions and costability conditions presented in
\cite{Ann}.

We want to calculate whether there exists a common eigenvector $v$ in the
costable condition \cite{math2}%
\begin{subequations}
\begin{align}
\left(  zB_{11}+wB_{21}\right)  v  &  =-xv,\label{3a}\\
\left(  zB_{12}+wB_{22}\right)  v  &  =-yv,\label{3b}\\
\left(  zJ_{1}+wJ_{2}\right)  v  &  =0 \label{3c}%
\end{align}
where $[x:y:z:w]$ are homogeneous coordinates of $CP^{3}$. If the common
eigenvector $v$ exists, then the dimension of the quotient space $($Ker
$\beta$/ Im $\alpha)$ in the monad construction will not be a constant since
the map $\alpha$ fail to be injective. In this case, the holomorphic vector
bundle description on $CP^{3}$ break down and one is led to use sheaf to
describe instanton on $CP^{3}$.

We first use Eq.(\ref{3c}) to obtain%
\end{subequations}
\begin{equation}
v\text{ }\sim%
\begin{bmatrix}
-ibc\left(  w^{2}-z^{2}\right) \\
-ac\left(  z^{2}+w^{2}\right) \\
-2iabzw
\end{bmatrix}
.
\end{equation}
On the other hand, Eq.(\ref{3a}) and Eq.(\ref{3b}) give
\begin{align}%
\begin{bmatrix}
x & -icz & -bw\\
-icz & x & -iaw\\
-bw & -iaw & x
\end{bmatrix}
v  &  =0,\label{3aa}\\%
\begin{bmatrix}
y & icw & bz\\
icw & y & -iaz\\
bz & -iaz & y
\end{bmatrix}
v  &  =0. \label{3bb}%
\end{align}

For simplicity, let's choose
\begin{equation}
z\in C,w=0 \label{zw}%
\end{equation}
on $CP^{3}$, then $v$ becomes%
\begin{equation}
v\text{ }\sim%
\begin{bmatrix}
ib\\
-a\\
0
\end{bmatrix}
. \label{vec}%
\end{equation}

The two characteristic equations corresponding to Eq.(\ref{3aa}) and
Eq.(\ref{3bb}) become%
\begin{align}
x\left(  x^{2}+c^{2}z^{2}\right)   &  =0,\label{33a}\\
y\left[  y^{2}+\left(  a^{2}-b^{2}\right)  z^{2}\right]   &  =0, \label{33b}%
\end{align}
which give the solutions%
\begin{align}
x  &  =0\text{ or }\pm icz,\\
y  &  =0\text{ or }\pm i\sqrt{a^{2}-b^{2}}z.
\end{align}

There are four cases for the choices of $x$ and $y$ above. For the first case
we choose $x=0$ and $y=0$, then with Eq.(\ref{vec}), Eq.(\ref{3aa}) and
Eq.(\ref{3bb}) become
\begin{align}%
\begin{bmatrix}
0 & -icz & 0\\
-icz & 0 & 0\\
0 & 0 & 0
\end{bmatrix}%
\begin{bmatrix}
ib\\
-a\\
0
\end{bmatrix}
&  =0,\\%
\begin{bmatrix}
0 & 0 & bz\\
0 & 0 & -iaz\\
bz & -iaz & 0
\end{bmatrix}%
\begin{bmatrix}
ib\\
-a\\
0
\end{bmatrix}
&  =0,
\end{align}
which give
\begin{equation}
c=0,a=\pm ib,b\neq0. \label{data1}%
\end{equation}

For the second case, we choose $x=\pm icz$ and $y=0$. In this case
Eq.(\ref{3aa}) and Eq.(\ref{3bb}) become%
\begin{align}%
\begin{bmatrix}
\pm icz & -icz & 0\\
-icz & \pm icz & 0\\
0 & 0 & \pm icz
\end{bmatrix}%
\begin{bmatrix}
ib\\
-a\\
0
\end{bmatrix}
&  =0,\\%
\begin{bmatrix}
0 & 0 & bz\\
0 & 0 & -iaz\\
bz & -iaz & 0
\end{bmatrix}%
\begin{bmatrix}
ib\\
-a\\
0
\end{bmatrix}
&  =0,
\end{align}
which give%
\begin{equation}
c\in C,a=\pm ib,b\neq0. \label{data2}%
\end{equation}

We conclude that for the ADHM data given at Eq.(\ref{data2}) and at the point
$[x:y:z:w]=[\pm ic:0:1:0]$ on $CP^{3}$, the map $\alpha$ fails to be
injective, thus one is led to use sheaf description for these (weakly)
instanton sheaves on $CP^{3}$ \cite{math2}. Note that for the $c=0$ case,
Eq.(\ref{data2}) reduces to Eq.(\ref{data1}).

It is important to note that for the case of $SU(2)$ $3$-instanton, $a$ and
$b$ are both real numbers which are inconsistent with Eq.(\ref{data2}). So the
corresponding $SU(2)$ $3$-instanton solutions are locally free. This is
consistent with the known vector bundle description of $SU(2)$ $3$-instanton
on $CP^{3}$.

For the third case, we choose $x=0$ and $y=\pm i\sqrt{a^{2}-b^{2}}z$. In this
case Eq.(\ref{3aa}) and Eq.(\ref{3bb}) become
\begin{align}%
\begin{bmatrix}
0 & -icz & 0\\
-icz & 0 & 0\\
0 & 0 & 0
\end{bmatrix}%
\begin{bmatrix}
ib\\
-a\\
0
\end{bmatrix}
&  =0,\\%
\begin{bmatrix}
\pm i\sqrt{a^{2}-b^{2}}z & 0 & bz\\
0 & \pm i\sqrt{a^{2}-b^{2}}z & -iaz\\
bz & -iaz & \pm i\sqrt{a^{2}-b^{2}}z
\end{bmatrix}%
\begin{bmatrix}
ib\\
-a\\
0
\end{bmatrix}
&  =0,
\end{align}
which give%
\begin{equation}
c=0,a=\pm ib,b\neq0. \label{data3}%
\end{equation}

For the fourth case, we choose $x=\pm icz$ and $y=\pm i\sqrt{a^{2}-b^{2}}z$.
In this case Eq.(\ref{3aa}) and Eq.(\ref{3bb}) become%
\begin{align}%
\begin{bmatrix}
\pm icz & -icz & 0\\
-icz & \pm icz & 0\\
0 & 0 & \pm icz
\end{bmatrix}%
\begin{bmatrix}
ib\\
-a\\
0
\end{bmatrix}
&  =0,\\%
\begin{bmatrix}
\pm i\sqrt{a^{2}-b^{2}}z & 0 & bz\\
0 & \pm i\sqrt{a^{2}-b^{2}}z & -iaz\\
bz & -iaz & \pm i\sqrt{a^{2}-b^{2}}z
\end{bmatrix}%
\begin{bmatrix}
ib\\
-a\\
0
\end{bmatrix}
&  =0,
\end{align}
which give%
\begin{equation}
c\in C,a=\pm ib,b\neq0. \label{data4}%
\end{equation}

We conclude that for the ADHM data given at Eq.(\ref{data4}) and at the point
$[x:y:z:w]=[\pm ic:\mp\sqrt{2}b:1:0]$ on $CP^{3}$, the map $\alpha$ fails to
be injective, thus one is led to use sheaf description for these (weakly)
instanton sheaves on $CP^{3}$. Note that for the $c=0$ case, Eq.(\ref{data4})
reduces to Eq.(\ref{data3}). Again $SU(2)$ instanton sheaf is not allowed for
this case.

Note that in the above $3$-instanton calculation, we have assumed $w=0$ on
$CP^{3}$ in Eq.(\ref{zw}). We expect that other choices of points on $CP^{3}$
will give more $3$-instanton sheaf structure for some other ADHM data. We
conclude the discussion of costable conditions of (weakly) $3$-instanton
sheaves. In the next subsection, we turn to discuss the stable conditions of
(weakly) $3$-instanton sheaves.

\subsection{The $\beta$ matrix and stable conditions}

To complete the description of (weakly) instanton sheaves, we need to check
the stable conditions \cite{math2}. For the ADHM data calculated in
Eq.(\ref{data2}), if there exist some points on $CP^{3}$ which satisfies the
stable conditions (no non-zero $v$ solution) \cite{math2}%
\begin{align}
\left(  \bar{z}B_{11}^{\dagger}+\bar{w}B_{21}^{\dagger}+\bar{x}\right)  v &
=0,\label{3d}\\
\left(  \bar{z}B_{12}^{\dagger}+\bar{w}B_{22}^{\dagger}+\bar{y}\right)  v &
=0,\label{3e}\\
\left(  \bar{z}I_{1}^{\dagger}+\bar{w}I_{2}^{\dagger}\right)  v &
=0,\label{3f}%
\end{align}
then we obtain the so-called weakly instanton sheaves \cite{math2}. In case
that the stable conditions are satisfied for all points on $CP^{3}$, then we
have the instanton sheaves. For simplicity, we will only check for the case of
$3$-instanton ADHM data given in Eq.(\ref{data4}) with $c=0$%
\begin{equation}
c=0,a=\pm ib,b\neq0.\label{data33}%
\end{equation}
For this case the weakly instanton sheaves are easy to check. In addition, we
will also show that there exist finitely many points \cite{math2} on $CP^{3}$
for which the stable conditions are not satisfied.

Since the $B$ matrices are symmetric, the stable conditions can be re-written
as%
\begin{align}
\left(  zB_{11}+wB_{21}+x\right)  \bar{v}  &  =0,\label{3dd}\\
\left(  zB_{12}+wB_{22}+y\right)  \bar{v}  &  =0,\label{3ee}\\
\left(  zI_{1}^{T}+wI_{2}^{T}\right)  \bar{v}  &  =0. \label{3ff}%
\end{align}
We are going to check whether there exist common non-zero vector $\bar{v}$ for
the stable conditions. Let's first work out the non-zero solutions of
Eq.(\ref{3ff}) which can be written as%
\begin{equation}%
\begin{pmatrix}
\mp bz & -bz & -icw\\
\pm bw & -bw & -icz
\end{pmatrix}%
\begin{pmatrix}
\bar{v}_{1}\\
\bar{v}_{2}\\
\bar{v}_{3}%
\end{pmatrix}
=0 \label{pp}%
\end{equation}
where we have used $a=\pm ib$ in Eq.(\ref{data33}). If one chooses $c=0$, the
solution is%
\begin{equation}%
\begin{pmatrix}
\bar{v}_{1}\\
\bar{v}_{2}\\
\bar{v}_{3}%
\end{pmatrix}
=%
\begin{pmatrix}
ibc\left(  z^{2}-w^{2}\right) \\
\mp ibc\left(  z^{2}+w^{2}\right) \\
\pm2b^{2}zw
\end{pmatrix}
=%
\begin{pmatrix}
0\\
0\\
\pm2b^{2}zw
\end{pmatrix}
\sim%
\begin{pmatrix}
0\\
0\\
1
\end{pmatrix}
\label{zz}%
\end{equation}
for all points on $CP^{3}$. It is important to note that for points on
$CP^{3}$ with $w=0$ and $z\neq0$, one gets additional solutions%
\begin{equation}%
\begin{pmatrix}
\bar{v}_{1}\\
\bar{v}_{2}\\
\bar{v}_{3}%
\end{pmatrix}
=%
\begin{pmatrix}
1\\
\mp1\\
0
\end{pmatrix}
. \label{(3)}%
\end{equation}
Similarly, for points on $CP^{3}$ with $z=0$ and $w\neq0$, one gets additional
solutions%
\begin{equation}%
\begin{pmatrix}
\bar{v}_{1}\\
\bar{v}_{2}\\
\bar{v}_{3}%
\end{pmatrix}
=%
\begin{pmatrix}
1\\
\pm1\\
0
\end{pmatrix}
. \label{(3')}%
\end{equation}
We have exhausted all solutions of Eq.(\ref{3ff}) for the ADHM data in
Eq.(\ref{data33}).

We next consider Eq.(\ref{3ee}) which can be written as (for $c=0$)%
\begin{equation}%
\begin{pmatrix}
y & 0 & bz\\
0 & y & -iaz\\
bz & -iaz & y
\end{pmatrix}%
\begin{pmatrix}
\bar{v}_{1}\\
\bar{v}_{2}\\
\bar{v}_{3}%
\end{pmatrix}
=0. \label{bb}%
\end{equation}
The characteristic equation can be calculated to be%
\begin{equation}
y^{3}-z^{2}\left(  b^{2}-a^{2}\right)  y=0,
\end{equation}
whose solutions are%
\begin{equation}
y=0,\pm z\sqrt{b^{2}-a^{2}}=\pm\sqrt{2}zb.
\end{equation}
For $y=0$ on $CP^{3}$, we get%
\begin{equation}%
\begin{pmatrix}
\bar{v}_{1}\\
\bar{v}_{2}\\
\bar{v}_{3}%
\end{pmatrix}
=%
\begin{pmatrix}
\mp1\\
1\\
0
\end{pmatrix}
. \label{(2)}%
\end{equation}
For $y=+\sqrt{2}zb$ on $CP^{3}$, we get%
\begin{equation}%
\begin{pmatrix}
\bar{v}_{1}\\
\bar{v}_{2}\\
\bar{v}_{3}%
\end{pmatrix}
=%
\begin{pmatrix}
1\\
\pm1\\
-\sqrt{2}%
\end{pmatrix}
.
\end{equation}
Finally, for $y=-\sqrt{2}zb$ on $CP^{3}$, we get%
\begin{equation}%
\begin{pmatrix}
\bar{v}_{1}\\
\bar{v}_{2}\\
\bar{v}_{3}%
\end{pmatrix}
=%
\begin{pmatrix}
1\\
\pm1\\
\sqrt{2}%
\end{pmatrix}
.
\end{equation}
It is important to note that for $z=0$ on $CP^{3}$, Eq.(\ref{bb}) can be
written as
\begin{equation}%
\begin{pmatrix}
y & 0 & 0\\
0 & y & 0\\
0 & 0 & y
\end{pmatrix}%
\begin{pmatrix}
\bar{v}_{1}\\
\bar{v}_{2}\\
\bar{v}_{3}%
\end{pmatrix}
=0
\end{equation}
The characteristic equation can be calculated to be%
\begin{equation}
y^{3}=0,
\end{equation}
whose solutions are%
\begin{equation}
y=0,\forall\bar{v}\in C^{3}. \label{(2')}%
\end{equation}
\qquad\qquad

Finally, we consider Eq.(\ref{3dd}) which can be written as (for $c=0$)%
\begin{equation}%
\begin{pmatrix}
x & 0 & -bw\\
0 & x & -iaw\\
-bw & -iaw & x
\end{pmatrix}%
\begin{pmatrix}
\bar{v}_{1}\\
\bar{v}_{2}\\
\bar{v}_{3}%
\end{pmatrix}
=0. \label{aa}%
\end{equation}

The characteristic equation can be calculated to be%
\begin{equation}
x^{3}-w^{2}\left(  b^{2}-a^{2}\right)  x=0,
\end{equation}
whose solutions are%
\begin{equation}
x=0,\pm w\sqrt{b^{2}-a^{2}}=\pm\sqrt{2}wb.
\end{equation}
For $x=0$ on $CP^{3}$, we get%
\begin{equation}%
\begin{pmatrix}
\bar{v}_{1}\\
\bar{v}_{2}\\
\bar{v}_{3}%
\end{pmatrix}
=%
\begin{pmatrix}
\pm1\\
1\\
0
\end{pmatrix}
. \label{(1)}%
\end{equation}
For $x=\sqrt{2}zb$ on $CP^{3}$, we get%
\begin{equation}%
\begin{pmatrix}
\bar{v}_{1}\\
\bar{v}_{2}\\
\bar{v}_{3}%
\end{pmatrix}
=%
\begin{pmatrix}
1\\
\mp1\\
\sqrt{2}%
\end{pmatrix}
.
\end{equation}
Finally, for $x=-\sqrt{2}zb$ on $CP^{3}$, we get%
\begin{equation}%
\begin{pmatrix}
\bar{v}_{1}\\
\bar{v}_{2}\\
\bar{v}_{3}%
\end{pmatrix}
=%
\begin{pmatrix}
1\\
\mp1\\
-\sqrt{2}%
\end{pmatrix}
.
\end{equation}
\bigskip It is important to note that for $w=0$ on $CP^{3}$, Eq.(\ref{aa}) can
be written as%
\begin{equation}%
\begin{pmatrix}
x & 0 & 0\\
0 & x & 0\\
0 & 0 & x
\end{pmatrix}%
\begin{pmatrix}
\bar{v}_{1}\\
\bar{v}_{2}\\
\bar{v}_{3}%
\end{pmatrix}
=0.
\end{equation}
The characteristic equation can be calculated to be%
\begin{equation}
x^{3}=0,
\end{equation}
whose solutions are%
\begin{equation}
x=0,\forall\bar{v}\in C^{3}. \label{(1')}%
\end{equation}
In summary, for the case of ADHM data with $a=+ib$ ($b\neq0,c=0$), we can make
the following choices of points on $CP^{3}$ and obtain the corresponding
non-zero $\bar{v}$
\begin{align}
Eq.(\ref{(3)})\text{,}Eq.(\ref{(2)})\text{,}Eq.(\ref{(1')});\text{ }w  &
=0,y=0,x=0;%
\begin{pmatrix}
1\\
-1\\
0
\end{pmatrix}
.\\
Eq.(\ref{(3')})\text{,}Eq.(\ref{(2')})\text{,}Eq.(\ref{(1)});\text{ }z  &
=0,y=0,x=0;%
\begin{pmatrix}
1\\
1\\
0
\end{pmatrix}
. \label{gg1}%
\end{align}
Similarly, for the case of ADHM data with $a=-ib$ ($b\neq0,c=0$), we can make
the following choices of points on $CP^{3}$ and obtain the corresponding
non-zero $\bar{v}$
\begin{align}
Eq.(\ref{(3)})\text{,}Eq.(\ref{(2)})\text{,}Eq.(\ref{(1')});\text{ }w  &
=0,y=0,x=0;%
\begin{pmatrix}
1\\
1\\
0
\end{pmatrix}
.\\
Eq.(\ref{(3')})\text{,}Eq.(\ref{(2')})\text{,}Eq.(\ref{(1)});\text{ }z  &
=0,y=0,x=0;%
\begin{pmatrix}
1\\
-1\\
0
\end{pmatrix}
. \label{gg2}%
\end{align}
Note that if we choose both $w=0$ and $z=0$ in Eq.(\ref{pp}), we are forced to
choose $x=0$ and $y=0$ which are not allowed.

To obtain the conditions of weakly instanton sheaves, one can choose say $w=1$
and $z=1$ in Eq.(\ref{zz}). It is easy to see that for all points on $CP^{3}$
(infinite number of) with $w=1$ and $z=1$, there exist no common non-zero
solution for the stable conditions.

\subsection{A Dual symmetry}

A dual symmetry between solutions of costable conditions and solutions of
stable conditions was found in \cite{LLT4}. To be more precisely, for an ADHM
data with given solutions of costable conditions, one can obtain solutions of
stable conditions with the new ADHM data%
\begin{equation}
(B_{11}^{\prime},B_{12}^{\prime},B_{21}^{\prime},B_{22}^{\prime},I_{1}%
^{\prime},I_{2}^{\prime},,J_{1}^{\prime},J_{2}^{\prime})=(B_{22}^{\dagger
},-B_{21}^{\dagger},-B_{12}^{\dagger},B_{11}^{\dagger},J_{2}^{\dagger}%
,-J_{1}^{\dagger},-I_{2}^{\dagger},I_{1}^{\dagger}),
\end{equation}
and at the new point%
\begin{equation}
\lbrack x^{\prime}:y^{\prime}:z^{\prime}:w^{\prime}]=[\bar{y}:-\bar{x}:\bar
{w}:-\bar{z}]
\end{equation}
on $CP^{3}$. For the case we are considering
\begin{align}
J_{1}  &  =%
\begin{pmatrix}
0 & 0 & 0\\
-ia & b & 0
\end{pmatrix}
,J_{2}=%
\begin{pmatrix}
-ia & -b & 0\\
0 & 0 & 0
\end{pmatrix}
,I_{1}=%
\begin{pmatrix}
ia & 0\\
-b & 0\\
0 & 0
\end{pmatrix}
,I_{2}=%
\begin{pmatrix}
0 & -ia\\
0 & -b\\
0 & 0
\end{pmatrix}
,\nonumber\\
B_{11}  &  =%
\begin{pmatrix}
0 & 0 & 0\\
0 & 0 & 0\\
0 & 0 & 0
\end{pmatrix}
,B_{21}=%
\begin{pmatrix}
0 & 0 & -b\\
0 & 0 & -ia\\
-b & -ia & 0
\end{pmatrix}
,B_{12}=%
\begin{pmatrix}
0 & 0 & b\\
0 & 0 & -ia\\
b & -ia & 0
\end{pmatrix}
,B_{22}=%
\begin{pmatrix}
0 & 0 & 0\\
0 & 0 & 0\\
0 & 0 & 0
\end{pmatrix}
,
\end{align}
and%
\begin{align}
J_{1}^{\prime}  &  =-I_{2}^{\dagger}=%
\begin{pmatrix}
0 & 0 & 0\\
-i\bar{a} & \bar{b} & 0
\end{pmatrix}
,J_{2}^{\prime}=I_{1}^{\dagger}=%
\begin{pmatrix}
-i\bar{a} & -\bar{b} & 0\\
0 & 0 & 0
\end{pmatrix}
,\nonumber\\
I_{1}^{\prime}  &  =J_{2}^{\dagger}=%
\begin{pmatrix}
i\bar{a} & 0\\
-\bar{b} & 0\\
0 & 0
\end{pmatrix}
,I_{2}^{\prime}=-J_{1}^{\dagger}=%
\begin{pmatrix}
0 & -i\bar{a}\\
0 & -\bar{b}\\
0 & 0
\end{pmatrix}
,\nonumber\\
B_{11}^{\prime}  &  =B_{22}^{\dagger}=%
\begin{pmatrix}
0 & 0 & 0\\
0 & 0 & 0\\
0 & 0 & 0
\end{pmatrix}
,B_{21}^{\prime}=-B_{12}^{\dagger}=%
\begin{pmatrix}
0 & 0 & -\bar{b}\\
0 & 0 & -i\bar{a}\\
-\bar{b} & -i\bar{a} & 0
\end{pmatrix}
,\nonumber\\
B_{12}^{\prime}  &  =-B_{21}^{\dagger}=%
\begin{pmatrix}
0 & 0 & \bar{b}\\
0 & 0 & -i\bar{a}\\
\bar{b} & -i\bar{a} & 0
\end{pmatrix}
,B_{22}^{\prime}=B_{11}^{\dagger}=%
\begin{pmatrix}
0 & 0 & 0\\
0 & 0 & 0\\
0 & 0 & 0
\end{pmatrix}
.
\end{align}
Note that for this case, the set of new ADHM data can be obtained by renaming
the set of old ADHM data by doing $a\rightarrow\bar{a},b\rightarrow\bar
{b},c=0\rightarrow c=0$. For example, for the case $a=+ib$, we have the dual
transformation of the ADHM data $(a,b,c)=(i,1,0)\rightarrow(-i,1,0)$. The old
and the new $CP^{3}$ points are%

\begin{align}
\lbrack x,y,z,w]  &  =[0,0,1,0],\nonumber\\
\lbrack x^{\prime},y^{\prime},z^{\prime},w^{\prime}]  &  =[\bar{y},-\bar
{x},\bar{w},-\bar{z}]=[0,0,0,-1].
\end{align}
We can see that the stable solutions of Eq.(\ref{gg1}) and Eq.(\ref{gg2}) can
be obtained from the solutions of costable solutions calculated previously.

\section{The Four (Weakly) Instanton Sheaves}

As has been well known, there are no complete $SU(2)$ $4$-instanton explicit
solutions in the literature. In \cite{4in}, a\textit{ two} parameter family of
$SU(2)$ $4$-instanton ADHM data was constructed \cite{4in}%
\begin{equation}
A_{4}=%
\begin{bmatrix}
\sqrt{2}be_{0} & \sqrt{2}be_{1} & \sqrt{2}be_{2} & \sqrt{2}be_{3}\\
a(e_{1}+e_{2}+e_{3}) & -b(e_{2}+e_{3}) & -b(e_{3}+e_{1}) & -b(e_{1}+e_{2})\\
-b(e_{2}+e_{3}) & a(e_{1}-e_{2}-e_{3}) & b(e_{2}-e_{1}) & b(e_{1}-e_{3})\\
-b(e_{3}+e_{1}) & b(e_{2}-e_{1}) & a(-e_{1}+e_{2}-e_{3}) & b(e_{3}-e_{2})\\
-b(e_{1}+e_{2}) & b(e_{1}-e_{3}) & b(e_{3}-e_{2}) & a(-e_{1}-e_{2}+e_{3})
\end{bmatrix}
, \label{44A}%
\end{equation}
which was used to calculate the approximate four Skyrme field. For this case,
it was shown that $A_{4}^{\ast}A_{4}=(8b^{2}+3a^{2})I_{4}$ , thus $A_{4}$ in
Eq.(\ref{44A}) does represent a class of $SU(2)$ ADHM $4$-instanton data
\cite{4in}.

For our purpose here, we want to study whether the corresponding $SL(2,C)$
ADHM $4$-instanton data contain the structure of (weakly) instanton sheaves or
not. To do the calculation, we first extend the parameters $a$ and $b$ in
Eq.(\ref{44A}) to complex numbers, and replace the quaternion calculation by
biquaternion calculation with biconjugation operation \cite{Ann}. One easily
gets $A_{4}^{\circledast}A_{4}=(8b^{2}+3a^{2})I_{4}$.

We then calculate $A_{4E}$ , the EMR of $A_{4}$, and do the rearrangement rule
to obtain $A_{4E}^{r}$ \cite{Ann2} and finally identify the corresponding ADHM
data $(B_{lm},I_{m,}J_{m})$ to be%
\begin{align}
J_{1}  &  =%
\begin{bmatrix}
\sqrt{2} & 0 & 0 & -i\sqrt{2}\\
0 & -i\sqrt{2} & \sqrt{2} & 0
\end{bmatrix}
,J_{2}=%
\begin{bmatrix}
0 & -i\sqrt{2} & -\sqrt{2} & 0\\
\sqrt{2} & 0 & 0 & i\sqrt{2}%
\end{bmatrix}
,\label{jj}\\
I_{1}  &  =%
\begin{pmatrix}
0 & \sqrt{2}\\
i\sqrt{2} & 0\\
-\sqrt{2} & 0\\
0 & -i\sqrt{2}%
\end{pmatrix}
,I_{2}=%
\begin{pmatrix}
-\sqrt{2} & 0\\
0 & -i\sqrt{2}\\
0 & -\sqrt{2}\\
-i\sqrt{2} & 0
\end{pmatrix}
\\
B_{11}  &  =%
\begin{bmatrix}
-ia & i & i & 0\\
i & ia & 0 & i\\
i & 0 & ia & -i\\
0 & i & -i & -ia
\end{bmatrix}
,B_{21}=%
\begin{bmatrix}
\left(  -1-i\right)  a & 1 & i & 1+i\\
1 & \left(  1-i\right)  a & -1+i & -i\\
i & -1+i & \left(  -1+i\right)  a & 1\\
1+i & -i & 1 & \left(  1+i\right)  a
\end{bmatrix}
,\\
B_{12}  &  =%
\begin{bmatrix}
\left(  1-i\right)  a & -1 & i & -1+i\\
-1 & \left(  -1-i\right)  a & 1+i & -i\\
i & 1+i & \left(  1+i\right)  a & -1\\
-1+i & -i & -1 & \left(  -1+i\right)  a
\end{bmatrix}
,B_{22}=%
\begin{bmatrix}
ia & -i & -i & 0\\
-i & -ia & 0 & -i\\
-i & 0 & -ia & i\\
0 & -i & i & ia
\end{bmatrix}
\end{align}
where we have put $b=1.$ The reason is as following. First we want to restrict
the two parameter $SL(2,C)$ ADHM data to be on $[a:b]$ $\in$ $CP^{1}$ and
simplify the calculation. Moreover, a general result in the mathematics
literature \cite{affine} claims to the effect that the moduli space
of\ unframed rank $2n$ instanton bundles over $CP^{2n+1}$ is an affine
variety. This result suggests that it is not the case that for all $[a:b]$ of
$CP^{1}$, the above ADHM data with parameters on $[a:b]$ (hence on $CP^{1}$)
gives only bundle solutions without exceptions, simply because it is well
known that an affine variety cannot contain any projective subvarieties of
positive dimension.

Indeed, we shall show below that for certain values of $[a:b]$, the above ADHM
data gives the sheaf (non-bundle) solutions. Our result will be consistent
with the mathematics result above.

\subsection{The $\alpha$ matrix and costable conditions}

The next step is to check whether there exists a common eigenvector $v$ in the
costable condition \cite{math2}. For simplicity, we choose $w=0$ and then put
$z=1$ on $CP^{3}$ without loss of generality
\begin{equation}
w=0,z=1. \label{wz}%
\end{equation}
For these choices, Eq.(\ref{3c}) gives two possible eigenvectors
\begin{equation}
v_{1}=%
\begin{bmatrix}
i\\
0\\
0\\
1
\end{bmatrix}
,v_{2}=%
\begin{bmatrix}
0\\
-i\\
1\\
0
\end{bmatrix}
.
\end{equation}
On the other hand, in order to have nontrivial $v$ solutions, Eq.(\ref{3a})
and Eq.(\ref{3b}) give the characteristic equations%
\begin{equation}
x^{4}+2x^{2}a^{2}+4x^{2}+a^{4}+4a^{2}+4=0,
\end{equation}%
\begin{equation}
y^{4}+4a^{4}+16a^{2}+12=0
\end{equation}
respectively. The solutions for $x$ and $y$ are%
\begin{equation}
x_{1}=\sqrt{-a^{2}-2},x_{2}=-\sqrt{-a^{2}-2}%
\end{equation}
and%
\begin{align}
y_{1}  &  =\left(  -4a^{4}-16a^{2}-12\right)  ^{\frac{1}{4}},y_{2}=i\left(
-4a^{4}-16a^{2}-12\right)  ^{\frac{1}{4}},\label{yy}\\
y_{3}  &  =-\left(  -4a^{4}-16a^{2}-12\right)  ^{\frac{1}{4}},y_{4}=-i\left(
-4a^{4}-16a^{2}-12\right)  ^{\frac{1}{4}}.\nonumber
\end{align}

We first choose $x=x_{1}=\sqrt{-a^{2}-2}$, then Eq.(\ref{3a}) becomes
\begin{equation}
\left(  B_{11}+\sqrt{-a^{2}-2}\right)  v=0,
\end{equation}
which gives two eigenvector solutions
\begin{equation}
v_{3}=%
\begin{bmatrix}
\frac{2i}{ia-\sqrt{-a^{2}-2}}\\
1\\
1\\
0
\end{bmatrix}
,v_{4}=%
\begin{bmatrix}
1\\
a+i\sqrt{-a^{2}-2}\\
0\\
1
\end{bmatrix}
.
\end{equation}
Since we need to have a common eigenvector to insure the instanton sheaf
structure, we impose the condition that the linear system%
\begin{equation}
c_{1}v_{1}+c_{2}v_{2}+c_{3}v_{3}+c_{4}v_{4}=0,
\end{equation}
or equivalently%
\begin{equation}%
\begin{bmatrix}
i & 0 & \frac{2i}{ia-\sqrt{-a^{2}-2}} & 1\\
0 & -i & 1 & a+i\sqrt{-a^{2}-2}\\
0 & 1 & 1 & 0\\
1 & 0 & 0 & 1
\end{bmatrix}%
\begin{bmatrix}
c_{1}\\
c_{2}\\
c_{3}\\
c_{4}%
\end{bmatrix}
=0, \label{cc}%
\end{equation}
contains nontrivial $(c_{1},c_{2},c_{3},c_{4})$ solutions. Surprisingly the
determinant of the coefficient matrix in Eq.(\ref{cc}) vanishes for
\textit{any} $a$ ! Thus one can easily calculate the solution%
\begin{equation}%
\begin{bmatrix}
c_{1}\\
c_{2}\\
c_{3}\\
c_{4}%
\end{bmatrix}
=%
\begin{bmatrix}
-1\\
\frac{-i}{2}\left(  1+i\right)  \left(  a+i\sqrt{-a^{2}-2}\right) \\
\frac{i}{2}\left(  1+i\right)  \left(  a+i\sqrt{-a^{2}-2}\right) \\
1
\end{bmatrix}
,
\end{equation}
which gives the common eigenvector
\begin{equation}
v=c_{1}v_{1}+c_{2}v_{2}=%
\begin{bmatrix}
-i\\
\left(  \frac{-1}{2}\right)  \left(  1+i\right)  \left(  a+i\sqrt{-a^{2}%
-2}\right) \\
\left(  \frac{-i}{2}\right)  \left(  1+i\right)  \left(  a+i\sqrt{-a^{2}%
-2}\right) \\
-1
\end{bmatrix}
. \label{v}%
\end{equation}
Finally we need to check whether the common eigenvector $v$ in Eq.(\ref{v})
satisfies Eq.(\ref{3b}).

\bigskip For the first choice of $y=y_{1}=\left(  -4a^{4}-16a^{2}-12\right)
^{\frac{1}{4}}$ in Eq.(\ref{yy}), Eq.(\ref{3b}) gives%
\begin{equation}
\left[  B_{12}+\left(  -4a^{4}-16a^{2}-12\right)  ^{\frac{1}{4}}I_{4}\right]
v=0,
\end{equation}
or explicitly%
\begin{align*}
-i\left[  \left(  1-i\right)  a+\left(  -4a^{4}-16a^{2}-12\right)  ^{\frac
{1}{4}}\right]  +\left(  1+i\right)  \left(  a+i\sqrt{-a^{2}-2}\right)  +1-i
&  =0,\\
2i+\left(  \frac{-1}{2}-\frac{i}{2}\right)  \left(  1+i\right)  \left[
\left(  -1-i\right)  a+\left(  -4a^{4}-16a^{2}-12\right)  ^{\frac{1}{4}%
}\right]  \left(  a+i\sqrt{-a^{2}-2}\right)  +\left(  a+i\sqrt{-a^{2}%
-2}\right)   &  =0,\\
2-i\left(  a+i\sqrt{-a^{2}-2}\right)  +\left(  \frac{1}{2}-\frac{i}{2}\right)
\left[  \left(  1+i\right)  a+\left(  -4a^{4}-16a^{2}-12\right)  ^{\frac{1}%
{4}}\right]  \left(  a+i\sqrt{-a^{2}-2}\right)   &  =0,\\
1+i+\left(  -1+i\right)  \left(  a+i\sqrt{-a^{2}-2}\right)  +\left(
1-i\right)  a-\left(  -4a^{4}-16a^{2}-12\right)  ^{\frac{1}{4}}  &  =0.
\end{align*}
The roots of these equations are%
\begin{align}
&  \sqrt{3}i,-\sqrt{3}i;\\
&  \sqrt{3}i,-\sqrt{3}i,\frac{1}{4}\sqrt{-26+2\sqrt{7}i},-\frac{1}{4}%
\sqrt{-26+2\sqrt{7}i};\nonumber\\
&  \sqrt{3}i,-\sqrt{3}i,\frac{1}{4}\sqrt{-26-2\sqrt{7}i},\frac{1}{4}%
\sqrt{-26-2\sqrt{7}i};\nonumber\\
&  \sqrt{3}i,-\sqrt{3}i;\nonumber
\end{align}
which again \textit{surprisingly} contain common roots%
\begin{equation}
a=\pm i\sqrt{3}.
\end{equation}

\bigskip We conclude that for the case of choosing
\begin{equation}
a=i\sqrt{3},
\end{equation}
the ADHM data are $J_{1}$ and $J_{2}$ in Eq.(\ref{jj}) and
\begin{align}
B_{11} &  =%
\begin{bmatrix}
\sqrt{3} & i & i & 0\\
i & -\sqrt{3} & 0 & i\\
i & 0 & -\sqrt{3} & -i\\
0 & i & -i & \sqrt{3}%
\end{bmatrix}
,B_{12}=%
\begin{bmatrix}
\sqrt{3}\left(  1+i\right)   & -1 & i & -1+i\\
-1 & \sqrt{3}\left(  1-i\right)   & 1+i & -i\\
i & 1+i & \sqrt{3}\left(  -1+i\right)   & -1\\
-1+i & -i & -1 & \sqrt{3}\left(  -1-i\right)
\end{bmatrix}
,\label{adhm11}\\
B_{21} &  =%
\begin{bmatrix}
\sqrt{3}\left(  1-i\right)   & 1 & i & 1+i\\
1 & \sqrt{3}\left(  1+i\right)   & -1+i & -i\\
i & -1+i & \sqrt{3}\left(  -1-i\right)   & 1\\
1+i & -i & 1 & \sqrt{3}\left(  -1+i\right)
\end{bmatrix}
,B_{22}=%
\begin{bmatrix}
-\sqrt{3} & -i & -i & 0\\
-i & \sqrt{3} & 0 & -i\\
-i & 0 & \sqrt{3} & i\\
0 & -i & i & -\sqrt{3}%
\end{bmatrix}
.\nonumber
\end{align}
There exists a common eigenvector of Eq.(\ref{3a}), Eq.(\ref{3b}) and
Eq.(\ref{3c})%
\begin{equation}
v=%
\begin{bmatrix}
-i\\
\frac{1}{2}\left(  1-i\right)  \left(  \sqrt{3}+1\right)  \\
\frac{i}{2}\left(  1-i\right)  \left(  \sqrt{3}+1\right)  \\
-1
\end{bmatrix}
.\label{v1}%
\end{equation}
The map $\alpha$ fails to be injective at%
\begin{equation}
\lbrack z:w:x:y]=[1:0:1:0]\label{cp3}%
\end{equation}
on $CP^{3}$, thus one is led to use sheaf description for this (weakly)
$4$-instanton sheaves on $CP^{3}$. Moreover, since $a$ is not a real number,
again $SU(2)$ instanton sheaf is not allowed for this case. This is consistent
with the common wisdom.

\bigskip Similarly, for the choice of
\begin{equation}
a=-i\sqrt{3},
\end{equation}
the ADHM data are $J_{1}$ and $J_{2}$ in Eq.(\ref{jj}) and%
\begin{align}
B_{11}  &  =%
\begin{bmatrix}
-\sqrt{3} & i & i & 0\\
i & \sqrt{3} & 0 & i\\
i & 0 & \sqrt{3} & -i\\
0 & i & -i & -\sqrt{3}%
\end{bmatrix}
,B_{12}=%
\begin{bmatrix}
\sqrt{3}\left(  -1-i\right)  & -1 & i & -1+i\\
-1 & \sqrt{3}\left(  -1+i\right)  & 1+i & -i\\
i & 1+i & \sqrt{3}\left(  1-i\right)  & -1\\
-1+i & -i & -1 & \sqrt{3}\left(  1+i\right)
\end{bmatrix}
,\label{adhm22}\\
B_{21}  &  =%
\begin{bmatrix}
\sqrt{3}\left(  -1+i\right)  & 1 & i & 1+i\\
1 & \sqrt{3}\left(  -1-i\right)  & -1+i & -i\\
i & -1+i & \sqrt{3}\left(  1+i\right)  & 1\\
1+i & -i & 1 & \sqrt{3}\left(  1-i\right)
\end{bmatrix}
,B_{22}=%
\begin{bmatrix}
\sqrt{3} & -i & -i & 0\\
-i & -\sqrt{3} & 0 & -i\\
-i & 0 & -\sqrt{3} & i\\
0 & -i & i & \sqrt{3}%
\end{bmatrix}
,\nonumber
\end{align}
and the common eigenvector is%
\begin{equation}
v=%
\begin{bmatrix}
-i\\
\frac{1}{2}\left(  1+i\right)  \left(  \sqrt{3}-1\right) \\
\frac{i}{2}\left(  -1+i\right)  \left(  \sqrt{3}-1\right) \\
-1
\end{bmatrix}
. \label{v2}%
\end{equation}
The map $\alpha$ fails to be injective at the same point in Eq.(\ref{cp3}) on
$CP^{3}$, and one ends up with another instanton sheaf case for these ADHM data.

For the second choice of $y=y_{2}=i\left(  -4a^{4}-16a^{2}-12\right)
^{\frac{1}{4}}$ in Eq.(\ref{yy}), Eq.(\ref{3b}) gives%
\begin{equation}
\left[  B_{12}+i\left(  -4a^{4}-16a^{2}-12\right)  ^{\frac{1}{4}}I_{4}\right]
v=0
\end{equation}
or explicitly%
\begin{align*}
-i\left[  \left(  1-i\right)  a+i\left(  -4a^{4}-16a^{2}-12\right)  ^{\frac
{1}{4}}\right]  +\left(  1+i\right)  \left(  a+i\sqrt{-a^{2}-2}\right)  +1-i
&  =0,\\
y\left(  a+i\sqrt{-a^{2}-2}\right)  +\left(  a+i\sqrt{-a^{2}-2}\right)   &
=0,\\
2-i\left(  a+i\sqrt{-a^{2}-2}\right)  +\left(  \frac{1}{2}-\frac{i}{2}\right)
\left[  \left(  1+i\right)  a+i\left(  -4a^{4}-16a^{2}-12\right)  ^{\frac
{1}{4}}\right]  \left(  a+i\sqrt{-a^{2}-2}\right)   &  =0,\\
1+i+\left(  -1+i\right)  \left(  a+i\sqrt{-a^{2}-2}\right)  +\left(
1-i\right)  a-i\left(  -4a^{4}-16a^{2}-12\right)  ^{\frac{1}{4}}  &  =0.
\end{align*}
The roots of these equations are%
\begin{align}
&  \sqrt{3}i,-\sqrt{3}i;\\
&  \sqrt{3}i,-\sqrt{3}i;\nonumber\\
&  \sqrt{3}i,-\sqrt{3}i,\frac{1}{4}\sqrt{-26-2\sqrt{7}i},\frac{1}{4}%
\sqrt{-26-2\sqrt{7}i};\nonumber\\
&  \sqrt{3}i,-\sqrt{3}i,\frac{1}{4}\sqrt{-26+2\sqrt{7}i},-\frac{1}{4}%
\sqrt{-26+2\sqrt{7}i};\nonumber
\end{align}
which again contain common roots
\begin{equation}
a=\pm i\sqrt{3}.
\end{equation}

For the third choice of $y=y_{3}=-\left(  -4a^{4}-16a^{2}-12\right)
^{\frac{1}{4}}$ in Eq.(\ref{yy}), Eq.(\ref{3b}) gives%
\begin{equation}
\left[  B_{12}-\left(  -4a^{4}-16a^{2}-12\right)  ^{\frac{1}{4}}I_{4}\right]
v=0,
\end{equation}
or explicitly%
\begin{align*}
-i\left[  \left(  1-i\right)  a-\left(  -4a^{4}-16a^{2}-12\right)  ^{\frac
{1}{4}}\right]  +\left(  1+i\right)  \left(  a+i\sqrt{-a^{2}-2}\right)  +1-i
&  =0,\\
2i+\left(  \frac{-1}{2}-\frac{i}{2}\right)  \left(  1+i\right)  \left[
\left(  -1-i\right)  a-\left(  -4a^{4}-16a^{2}-12\right)  ^{\frac{1}{4}%
}\right]  \left(  a+i\sqrt{-a^{2}-2}\right)  +\left(  a+i\sqrt{-a^{2}%
-2}\right)   &  =0,\\
2-i\left(  a+i\sqrt{-a^{2}-2}\right)  +\left(  \frac{1}{2}-\frac{i}{2}\right)
\left[  \left(  1+i\right)  a-\left(  -4a^{4}-16a^{2}-12\right)  ^{\frac{1}%
{4}}\right]  \left(  a+i\sqrt{-a^{2}-2}\right)   &  =0,\\
1+i+\left(  -1+i\right)  \left(  a+i\sqrt{-a^{2}-2}\right)  +\left(
1-i\right)  a+\left(  -4a^{4}-16a^{2}-12\right)  ^{\frac{1}{4}}  &  =0.
\end{align*}
The roots of these equations are%
\begin{align}
&  \sqrt{3}i,-\sqrt{3}i,\frac{1}{4}\sqrt{-26-2\sqrt{7}i},\frac{1}{4}%
\sqrt{-26-2\sqrt{7}i};\\
&  \sqrt{3}i,-\sqrt{3}i;\nonumber\\
&  \sqrt{3}i,-\sqrt{3}i,\frac{1}{4}\sqrt{-26+2\sqrt{7}i},-\frac{1}{4}%
\sqrt{-26+2\sqrt{7}i};\nonumber\\
&  \sqrt{3}i,-\sqrt{3}i;\nonumber
\end{align}
which again contain common roots
\begin{equation}
a=\pm i\sqrt{3}%
\end{equation}

For the fourth choice of $y=y_{4}=-i\left(  -4a^{4}-16a^{2}-12\right)
^{\frac{1}{4}}$ in Eq.(\ref{yy}), Eq.(\ref{3b}) gives%
\begin{equation}
\left[  B_{12}-i\left(  -4a^{4}-16a^{2}-12\right)  ^{\frac{1}{4}}I_{4}\right]
v=0
\end{equation}
or explicitly%
\begin{align*}
-i\left[  \left(  1-i\right)  a-i\left(  -4a^{4}-16a^{2}-12\right)  ^{\frac
{1}{4}}\right]  +\left(  1+i\right)  \left(  a+i\sqrt{-a^{2}-2}\right)  +1-i
&  =0,\\
2i+\left(  \frac{-1}{2}-\frac{i}{2}\right)  \left(  1+i\right)  \left[
\left(  -1-i\right)  a-i\left(  -4a^{4}-16a^{2}-12\right)  ^{\frac{1}{4}%
}\right]  \left(  a+i\sqrt{-a^{2}-2}\right)  +\left(  a+i\sqrt{-a^{2}%
-2}\right)   &  =0,\\
2-i\left(  a+i\sqrt{-a^{2}-2}\right)  +\left(  \frac{1}{2}-\frac{i}{2}\right)
\left[  \left(  1+i\right)  a-i\left(  -4a^{4}-16a^{2}-12\right)  ^{\frac
{1}{4}}\right]  \left(  a+i\sqrt{-a^{2}-2}\right)   &  =0,\\
1+i+\left(  -1+i\right)  \left(  a+i\sqrt{-a^{2}-2}\right)  +\left(
1-i\right)  a+i\left(  -4a^{4}-16a^{2}-12\right)  ^{\frac{1}{4}}  &  =0.
\end{align*}
The roots of these equations are%
\begin{align}
&  \sqrt{3}i,-\sqrt{3}i,\frac{1}{4}\sqrt{-26+2\sqrt{7}i},-\frac{1}{4}%
\sqrt{-26+2\sqrt{7}i};\\
&  \sqrt{3}i,-\sqrt{3}i,\frac{1}{4}\sqrt{-26-2\sqrt{7}i},\frac{1}{4}%
\sqrt{-26-2\sqrt{7}i};\nonumber\\
&  \sqrt{3}i,-\sqrt{3}i;\nonumber\\
&  \sqrt{3}i,-\sqrt{3}i;\nonumber
\end{align}
which again contain common roots
\begin{equation}
a=\pm i\sqrt{3}. \label{sheaf}%
\end{equation}

We conclude that for $x=x_{1}=\sqrt{-a^{2}-2}$ and all four choices of $y$ in
Eq.(\ref{yy}), the map $\alpha$ fails to be injective at the same point
$[z:w:x:y]=[1:0:1:0]$ on $CP^{3}$ for the ADHM data in Eq.(\ref{jj}),
Eq.(\ref{adhm11}) and Eq.(\ref{adhm22}), and one is led to use sheaf
description for these $4$-instantons. Again $SU(2)$ instanton sheaf is not
allowed for all cases.

Finally, a moment of thought leads one to extend the above $4$-instanton sheaf
structure in Eq.(\ref{sheaf}) to more general ADHM data. We first note that
for $b\neq0$, Eq.(\ref{44A}) can be rewritten as%
\begin{equation}
A_{4}=b%
\begin{bmatrix}
\sqrt{2}e_{0} & \sqrt{2}e_{1} & \sqrt{2}e_{2} & \sqrt{2}e_{3}\\
\frac{a}{b}(e_{1}+e_{2}+e_{3}) & -(e_{2}+e_{3}) & -(e_{3}+e_{1}) &
-(e_{1}+e_{2})\\
-(e_{2}+e_{3}) & \frac{a}{b}(e_{1}-e_{2}-e_{3}) & (e_{2}-e_{1}) & (e_{1}%
-e_{3})\\
-(e_{3}+e_{1}) & (e_{2}-e_{1}) & \frac{a}{b}(-e_{1}+e_{2}-e_{3}) &
(e_{3}-e_{2})\\
-(e_{1}+e_{2}) & (e_{1}-e_{3}) & (e_{3}-e_{2}) & \frac{a}{b}(-e_{1}%
-e_{2}+e_{3})
\end{bmatrix}
.
\end{equation}
\bigskip Our previous results imply that for the ADHM data $b=1$ and $\frac
{a}{b}=\pm i\sqrt{3}$, there exist sheaf structure. Now we can easily show
that for the ADHM data%
\begin{equation}
a=\pm i\sqrt{3}b\text{, \ }b\neq0\text{, }b\in C,
\end{equation}
there are sheaf structure for the $4$-instanton. Indeed the same common
eigenvectors in Eq.(\ref{v1}), Eq.(\ref{v2}) exist for these ADHM data at
points%
\begin{equation}
\lbrack z:w:x:y]=[1:0:b:0]
\end{equation}
on $CP^{3}$. This result can be easily checked by using Eq.(\ref{3a}),
Eq.(\ref{3b}) and Eq.(\ref{3c}).

We have checked that for the second choice of $x=x_{2}=-\sqrt{-a^{2}-2}$,
there is no common eigenvector $v$ for the system and thus no sheaf structure
of the YM $4$-instantons.

In the above long calculation of searching YM (weakly) $4$-instanton sheaves,
we have assumed $w=0$ on $CP^{3}$ in Eq.(\ref{wz}). We expect that other
choices of points on $CP^{3}$ will give more (weakly) $4$-instanton sheaf
structure for some other ADHM data.

\subsection{The $\beta$ matrix and stable conditions}

To complete the description of (weakly) $4$-instanton sheaves, again we need
to check the stable conditions. We first check Eq.(\ref{3ff}) which can be
written as
\begin{equation}%
\begin{pmatrix}
-w & iz & -z & -iw\\
z & -iw & -w & -iz
\end{pmatrix}
\bar{v}=0.\label{33}%
\end{equation}
For simplicity, we can use the following two set of three row vectors%
\begin{align}
&
\begin{pmatrix}
-w & iz & -z & -iw
\end{pmatrix}
,\\
&
\begin{pmatrix}
z & -iw & -w & -iz
\end{pmatrix}
,\\
&
\begin{pmatrix}
1 & 0 & 0 & 0
\end{pmatrix}
;
\end{align}%
\begin{align}
&
\begin{pmatrix}
-w & iz & -z & -iw
\end{pmatrix}
,\\
&
\begin{pmatrix}
z & -iw & -w & -iz
\end{pmatrix}
,\\
&
\begin{pmatrix}
0 & 1 & 0 & 0
\end{pmatrix}
\end{align}
and wedge product to generate the two null vectors%
\begin{align}
\bar{v}_{1} &  =%
\begin{pmatrix}%
\begin{vmatrix}
iz & -z & -iw\\
-iw & -w & -iz\\
0 & 0 & 0
\end{vmatrix}
& -%
\begin{vmatrix}
-w & -z & -iw\\
z & -w & -iz\\
1 & 0 & 0
\end{vmatrix}
&
\begin{vmatrix}
-w & iz & -iw\\
z & -iw & -iz\\
1 & 0 & 0
\end{vmatrix}
& -%
\begin{vmatrix}
-w & iz & -z\\
z & -iw & -w\\
1 & 0 & 0
\end{vmatrix}
\end{pmatrix}
^{T}\nonumber\\
&  =%
\begin{pmatrix}
0 & -iz^{2}+iw^{2} & z^{2}+w^{2} & 2izw
\end{pmatrix}
^{T},\label{22}\\
\bar{v}_{2} &  =%
\begin{pmatrix}%
\begin{vmatrix}
iz & -z & -iw\\
-iw & -w & -iz\\
1 & 0 & 0
\end{vmatrix}
& -%
\begin{vmatrix}
-w & -z & -iw\\
z & -w & -iz\\
0 & 0 & 0
\end{vmatrix}
&
\begin{vmatrix}
-w & iz & -iw\\
z & -iw & -iz\\
0 & 1 & 0
\end{vmatrix}
& -%
\begin{vmatrix}
-w & iz & -z\\
z & -iw & -w\\
0 & 1 & 0
\end{vmatrix}
\end{pmatrix}
^{T}\nonumber\\
&  =%
\begin{pmatrix}
iz^{2}-iw^{2} & 0 & -2izw & z^{2}+w^{2}%
\end{pmatrix}
^{T}.\label{11}%
\end{align}
Any linear combination of vectors in Eq.(\ref{22}) and Eq.(\ref{11}) is a
solution of Eq.(\ref{33}). We will show that for the ADHM data given in
Eq.(\ref{sheaf}), there exists some point on $CP^{3}$ where the stable
conditions are not satisfied. If we take $z=1,w=0$, then
\begin{equation}
\bar{v}_{1}=%
\begin{pmatrix}
0\\
-i\\
1\\
0
\end{pmatrix}
,\bar{v}_{2}=%
\begin{pmatrix}
i\\
0\\
0\\
1
\end{pmatrix}
.
\end{equation}
For the first choice of ADHM data $a=-\sqrt{3}i$ and taking $z=1,w=0$, the
second stable condition Eq.(\ref{3ee}) can be writtwn as%
\begin{equation}%
\begin{pmatrix}
-\sqrt{3}+x & i & i & 0\\
i & \sqrt{3}+x & 0 & i\\
i & 0 & \sqrt{3}+x & -i\\
0 & i & -i & -\sqrt{3}+x
\end{pmatrix}
\bar{v}=0.
\end{equation}
The characteristic equation%
\begin{equation}
x^{4}-2x^{2}+1=0
\end{equation}
gives two solutions%
\begin{equation}
x=+1,-1.
\end{equation}
For the case $x=1$, we get two degenerate null vectors%
\begin{equation}
\bar{v}_{3}=%
\begin{pmatrix}
1\\
-i\left(  \sqrt{3}-1\right)  \\
0\\
1
\end{pmatrix}
,\bar{v}_{4}=%
\begin{pmatrix}
i\left(  \sqrt{3}+1\right)  \\
1\\
1\\
0
\end{pmatrix}
.
\end{equation}
For the case $x=-1$, we also get two degenerate null vectors%
\begin{equation}
\bar{v}_{5}=%
\begin{pmatrix}
1\\
-i\left(  \sqrt{3}+1\right)  \\
0\\
1
\end{pmatrix}
,\bar{v}_{6}=%
\begin{pmatrix}
i\left(  \sqrt{3}-1\right)  \\
1\\
1\\
0
\end{pmatrix}
.
\end{equation}
For the first choice of ADHM data $a=-\sqrt{3}i$ and taking $z=1,w=0$, the
third stable condition Eq.(\ref{3dd}) can be writtwn as%
\begin{equation}%
\begin{pmatrix}
\left(  -1-i\right)  \sqrt{3}+y & -1 & i & -1+i\\
-1 & \left(  -1+i\right)  \sqrt{3}+y & 1+i & -i\\
i & 1+i & \left(  1-i\right)  \sqrt{3}+y & -1\\
-1+i & -i & -1 & \left(  1+i\right)  \sqrt{3}+y
\end{pmatrix}
\bar{v}=0.
\end{equation}
The characteristic equation is very simple%
\begin{equation}
y^{4}=0
\end{equation}
with solution $y=0$. One can work out the only non-zero solution to be
\begin{equation}
\bar{v}=%
\begin{pmatrix}
i\\
\frac{\left(  1-i\right)  \left(  \sqrt{3}-1\right)  }{2}\\
\frac{\left(  1+i\right)  \left(  \sqrt{3}-1\right)  }{2}\\
1
\end{pmatrix}
.
\end{equation}
Finally, one can check that $\bar{v}$ can be written as
\begin{align}
\bar{v} &  =\frac{\left(  1+i\right)  \left(  \sqrt{3}-1\right)  }{2}\bar
{v}_{1}+\bar{v}_{2}\\
&  =\bar{v}_{3}+\frac{\left(  1+i\right)  \left(  \sqrt{3}-1\right)  }{2}%
\bar{v}_{4}.
\end{align}
But $\bar{v}$ can not be written as linear combination of $\bar{v}_{5}$ and
$\bar{v}_{6}$. So $x=-1$ is not allowed. We conclude that there is a jumping
for $\beta$ at point%
\begin{equation}
\lbrack z:w:x:y]=[1:0:1:0]\label{point}%
\end{equation}
on $CP^{3}$ for which the stable conditions are not satisfied.

For the second choice of ADHM data $a=+\sqrt{3}i$, all the above calculations
for $a=-\sqrt{3}i$ go through with the replacement $\sqrt{3}\rightarrow
-\sqrt{3}$. So there is a jumping for $\beta$ at point in Eq.(\ref{point}).

To obtain the conditions of weakly instanton sheaves, one can choose say
$x=-1$. It is easy to see that at point%
\begin{equation}
\lbrack z:w:x:y]=[1:0:-1:0],
\end{equation}
the stable conditions are satisfied and there exists no common non-zero
solution for the stable conditions.

\section{Conclusion}

In this paper, we first construct a class of $SL(2,C)$ Yang-Mills ADHM
$3$-instanton data, and then demonstrate the existence of YM (weakly)
$3$-instanton sheaves on $CP^{3}$. We then use a class of two parameter ADHM
\textit{symmetric} $4$-instanton data constructed in the literature \cite{4in}
to demonstrate the existence of YM (weakly) $4$-instanton sheaves. The results
we obtained in this paper extend the recent construction of Yang-Mills
(weakly) $2$-instanton sheaves \cite{Ann2} to higher (weakly) instanton
sheaves. It is of interest to understand the relationship between YM
\textit{symmetric} instantons \cite{3in,4in,7in} on $S^{4}$ and YM instanton
\textit{sheaves} on $CP^{3}$ constructed in this paper.

Since it is a nontrivial task to explicitly construct non-diagonal \cite{Ann}
higher ADHM instanton data \cite{7in}, the explicit construction of the
general higher (weakly) $k$-instanton sheaves remains an open question.
However, it is believed that this new YM (weakly) instanton sheaf structure
persists for arbitrary higher $k$-instanton, and is a common feature for
non-compact SDYM theory which does not exist for the usual YM theory based on
the compact Lie group.

\begin{acknowledgments}
The work of J.C. is supported in part by the Ministry of Science and
Technology and S.T. Yau center of NCTU, Taiwan. The work of I-H. has been
possible due to an opportunity for him to visit S.T. Yau center of NCTU to
which he owes his thanks. We are grateful to the referee for his constructive
suggestions which lead to many improvements in Section III. B, C and IV. B of
this paper.
\end{acknowledgments}

\end{document}